\documentclass[
aps,%
12pt,%
final,%
notitlepage,%
oneside,%
onecolumn,%
nobibnotes,%
nofootinbib,%
superscriptaddress,%
noshowpacs,%
centertags]%
{revtex4}
\begin{document}


\title{DENSITY PROFILES IN MOLECULAR CLOUD CORES
ASSOCIATED WITH HIGH-MASS STAR-FORMING REGIONS}

\author{\firstname{L.~….}~\surname{Pirogov}}
\email{pirogov@appl.sci-nnov.ru}
\affiliation{%
Institute of Applied Physics RAS
}%

\begin{abstract}

Radial density profiles for the sample of dense cores
associated with high-mass star-forming regions from southern hemisphere
have been derived using the data of observations in continuum at 250~GHz.
Radial density profiles for the inner regions of 16 cores
(at distances $\la 0.2-0.8$~pc from the center) are close
on average to the $\rho\propto r^{-\alpha}$ dependence,
where $\alpha=1.6\pm 0.3$.
In the outer regions density drops steeper.
An analysis with various hydrostatic models showed
that the modified Bonnor-Ebert model, which describes turbulent sphere
confined by external pressure,
is preferable compared with the logotrope and polytrope models
practically in all cases.
With a help of the Bonnor-Ebert model, estimates of central density in a core,
non-thermal velocity dispersion and core size are obtained.
The comparison of central densities with the densities derived
earlier from the CS modeling reveals differences in several cases.
The reasons of such differences are probably connected
with the presence of density inhomogenities on the scales smaller
than the telescope beam.
In most cases non-thermal velocity dispersions are in agreement
with the values obtained from molecular line observations.


\end{abstract}

\maketitle

\section{Introduction}

The knowledge of density structure in star-forming cores
is very important for selecting most adequate theoretical models and
for understanding the processes which lead to formation of stars from dense gas.
Nowadays, there exists a general picture which describes the low-mass star
formation process ($M\sim 1~M_{\odot}$) (see, e.g. \cite{SAL,MO07}),
whereas the process of star formation in clusters, where high-mass stars
($M\ga 8~M_{\odot}$) are born, is studied much weaker.
The cores where low-mass stars are forming are assumed
to be initially in hydrostatic equilibrium in which gravitational force
is in balance with thermal and magnetic pressure.
During their evolution the cores loose stability and pass to the state of contraction.
For the case when inner pressure is purely thermal whereas central density
and radius of the sphere are finite, the stable solutions
of hydrostatic equations exist, if the sphere is confined by external pressure
(the Bonnor-Ebert model \cite{Bon56,Eb55}).
Recently, a number of papers appeared which confirm that this model
well describes the observed column density profiles in dense molecular cloud
cores where low-mass stars are forming
(see, e.g. \cite{ALL01,E01,Kan05}).
Most of the studied objects turn out to be either unstable
or near the critical state.
Model calculations show that the density distribution corresponding
to the Bonnor-Ebert model can remain for unstable objects \cite{M05,Kan05}.
Although this model describes isothermal sphere, whereas observations show
that there are non-thermal gas motions in the studied objects which lead
to additional line broadening, one can take them into account by adding
into equation of state the pressure due to non-thermal motions and
considering that it is uniform and isotropic (microturbulent approximation)
\cite{Lai03}.
However, the model described above is not without alternatives.
Recently, the gravoturbulent model of star formation,
where the observed cores can be unstable density fluctuations
in a turbulent cloud, is developed \cite{BP07}.
Nonetheless, density profiles in such cores are close
to those followed from the Bonnor-Ebert model \cite{BP03}.

The picture of high-mass star formation in dense cores, where non-thermal
turbulent motions prevail upon the thermal ones and evolution is going
considerably quicker than in low-mass star-forming regions, is still
under construction nowadays.
Several theoretical works suggest to use either polytrope models
($P\propto \rho^{1+1/N}$, where $P$ is a pressure, $\rho$ is a density)
or phenomenological ``logotrope" model \cite{MP96}
($P/P_c=1+A\ln(\rho/\rho_c)$, where
$P_c$ and $\rho_c$ are pressure and density in the center, respectively,
$A\approx 0.2$) as the equation of state for high-mass star-forming regions.
Both logotrope and polytrope equation with $N<-1$ predict
an increase of velocity dispersion from center to periphery corresponding
to the observed relations between line width and emission region size
(see, e.g. \cite{Lar81,CM95}).
However, a number of observations of high-mass star-forming regions shows that
non-thermal gas velocity dispersion is either constant or increases towards
the center \cite{CM95,Pir03} which also allows to consider polytropes
with positive $N$-index.
Note, that both polytrope spheres with $N>5$ and the spheres with $N<-1$
as well as the Bonnor-Ebert isothermal sphere model have infinite mass
and infinite radius, so the stable configurations can exist if
they are confined by a pressure from external medium \cite{VH74}.
In order for a sphere to be in equilibrium, a center-to-edge density ratio
should be not higher than 14.3 in the Bonnor-Ebert model
(see e.g. \cite{McCrea57}) while for the logotrope model the maximum value
for this ratio can reach $\ga 100$ \cite{MP96}.

Until now, the Bonnor-Ebert model is used in analysis of the observed gas
and dust distributions in the cores where low-mass stars are forming.
In the studies of density distributions in the cores associated with
high-mass star-forming regions one usually limits himself by the fitting
power-law dependences into observational data.
If object is in hydrostatic equilibrium, the power-law index of radial
density profile written in the form of power-law function
($\rho\propto r^{-\alpha}$) is close to zero in the center
and asymptotically tends to 2 (the Bonnor-Ebert model)
or to unity (the logotrope model) far from the center.
In the polytrope models the $\alpha$-index on periphery can accept
values between 1 and 2 depending on the value of $N$.
Both the logotrope and the polytrope models are not free of drawbacks
(see the critique in \cite{MP96,CM00,MT03}).
This leads to the necessity of considering more complex models \cite{CM00,MH99}.

In order to examine the models one needs experimental measurements
of density profiles which can be obtained from observations of
intensity distributions of optically thin molecular lines,
if radial excitation temperature profile is known,
or from dust continuum emission
if dust temperature profiles and dust absorption coefficient are known.
Recently, the methods based on estimates of star color variations
in near infrared due to absorption in dust cloud are used to estimate
dust column densities \cite{LAL99}.
In the present work an analysis of dust continuum maps obtained
towards several high-mass star-forming regions \cite{Pir07} is carried out.
The density profiles in the cores contained in these regions
have been estimated with a help of different models.
The results of the analysis are given below.

\section{Source selection}

The data of observations of the sample of high-mass star-forming
regions from southern hemisphere in continuum at 250~GHz \cite{Pir07}
have been used to estimate density profiles.
The source list with coordinates and distances is given in Table~\ref{list}.

\begin{figure}[t!]
\setcaptionmargin{5mm}
\onelinecaptionsfalse
\centering \includegraphics[width=15cm]{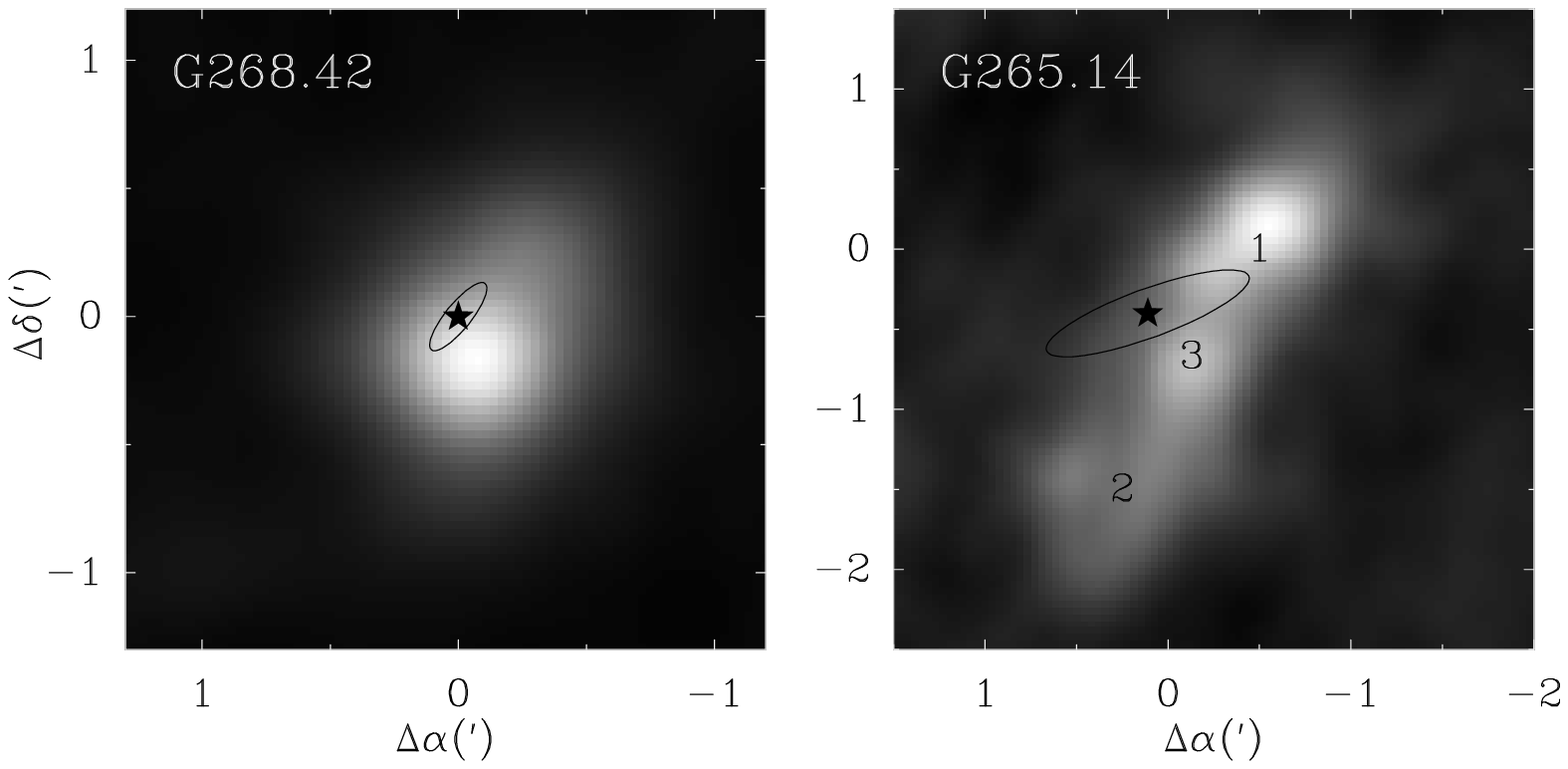}
\captionstyle{flushleft}
\caption{
The examples of continuum maps at 250~GHz taken from \cite{Pir07}
for two sample objects.
The positions of IRAS point sources are indicated by stars.
The uncertainties of these positions are indicated by ellipses.
Distinct cores for G265.14 are indicated by numbers.
}
\label{maps}
\end{figure}

The observed regions contain one or several intensity peaks which
can be associated with dense cores \cite{Pir07}.
Two examples of maps of the regions consisted of one and three cores,
respectively, are shown in Fig.~\ref{maps}.
Parameters of distinct cores are determined by fitting single or multiple
Gaussian elliptic distributions convolved with the telescope beam
(the values of the axes ratio of the fitted ellipse and the relative coordinates
of its center are given in Table~4 from \cite{Pir07}).
As far as the following analysis is performed within the framework
of spherically-symmetric model, we have selected 16 cores
which emission regions geometry is close
to the circular one (the axes ratio of the fitted ellipse does not exceed 2),
with total fluxes higher than 5~Jy and with sizes higher
or of the order of the main telescope beam (24$''$) \cite{Pir07}.
The iterative Marquard-Levenberg method has been used (see e.g. \cite{Press92})
for fitting model distributions convolved with the beam into the observed maps.
The values of model parameters and their uncertainties
have been determined on its basis.

\section{The analysis with power-law functions}
\label{powerlaw}

As far as the sources are optically thin at the frequency of observations
($\nu$), the intensity of their emission can be given by the following
integral along the line of sight (see e.g. \cite{Adams91}):

\begin{equation}
I_{\nu}=\int B_{\nu}[T(l)]\,\rho(l)\,k_{\nu}\,dl
\end{equation}

\noindent{where $B_{\nu}[T(l)]$ is the Plank function,
$T(l)$ is a dust temperature,
$\rho(l)$ is a density, $k_{\nu}$ is a dust absorption coefficient.
}
The convolution of $I_{\nu}$ with the telescope beam gives the observed
intensities.
If dust temperature along the line of sight is constant and the properties
of dust grains are also constant ($k_{\nu}=const$), the observed intensities
are proportional to dust and gas column densities
(given that the gas-to-dust mass ratio is constant)
in the column having cross-sectional area equal
to the projection of the main beam.

\begin{figure}[t!]
\setcaptionmargin{5mm}
\onelinecaptionsfalse
\centering \includegraphics[width=15cm]{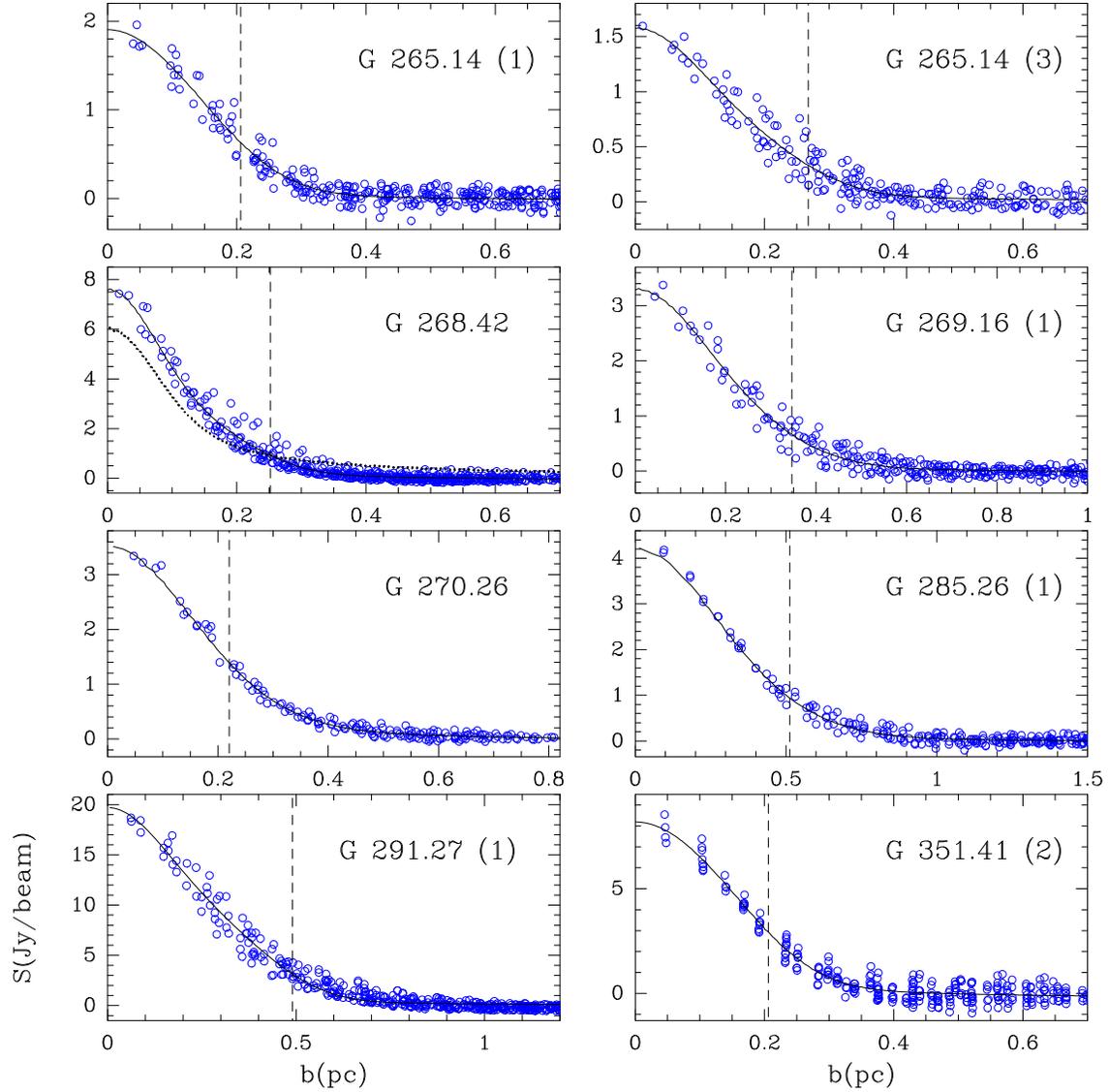}
\captionstyle{flushleft}
\caption{
The examples of the observed intensities and fitted two-dimensional
model distributions, represented as one-dimensional dependences
of flux densities on the distance from the center for eight sample objects.
The model distribution is a convolution of composite power-law function
and the telescope beam.
For G268.42 the result of fitting by a power-law function with single power-law
index is shown by dots (an additive constant term is set to zero).
Dashed vertical lines denote division radius between inner and outer regions
for which power-law indices are different.
}
\label{power-law}
\end{figure}

If one approximates density profile in the cores by power-law function,
it is possible
to estimate its power-law index from observational data.
For this purpose a two-dimensional power-law function with arbitrary power-law
index ($b^{-p}$, where $b$ is a projected distance from the center of map)
convolved with the beam has been fitted into the maps of the cores.
An amplitude of the fitting function, a value of additive constant term,
coordinates of a center and power-law index $p$ are varied
in order to get minimum of discrepancy.
As far as a function with single power-law index $p$ leads
to unsatisfactory results in most cases
(below, in Fig.~\ref{power-law} the results of fitting by such
a function is shown by dots for G268.42), it has been replaced by composite
power-law function for which power-law indices in inner and outer
core regions are different.
The value of division radius between these regions is varied
to get minimum of discrepancy.
The values of power-law indices for inner and outer regions
($p_i$ and $p_{out}$, respectively) as well as angular and linear values
of division radius between these regions are given in Table~\ref{powlaw}.
Uncertainties derived from the fitting are given in brackets.
The cores that belong to the same region are marked by numbers.
Their numbering is taken from \cite{Pir07}.
In addition, G291.27(4) with relative coordinates (182$''$,\,34$''$)
is included into the list.
For two cores in G316.77 and for G345.01(2) the results of fitting
by power-law function with single power-law index
for the whole core and convolved with the telescope beam are given
(the $p_i$ and $p_{out}$ values are the same for them).
A presence or an absence of the IRAS source within the core
is indicated in the last column of the table.
The examples of the observed intensities and fitted functions
for eight sample cores are shown in Fig.~\ref{power-law}
as one-dimensional dependences on the distance from core center.

If dust properties do not vary with radius at least for inner regions
of the cores, the power-law index for radial density profile
can be obtained from the following relation \cite{Adams91}:

\begin{equation}
\alpha\approx p+1-Q\,q
\end{equation}

\noindent{where $q$ is a power-law index for dust temperature radial dependence;
$Q=(x\,e^x)/(e^x-1)$ is a correction factor; $x=h\,\nu/K_{\rm B}\,T$\,;
$h$ and $K_{\rm B}$ are the Plank and the Bolzmann constants, respectively;
$\nu$ is a frequency; $T$ is a dust temperature.}
The $q$ value depends on $\beta$, the power-law index of the dependence
of dust optical depth on frequency ($q=2/(\beta+4)$ \cite{DL94}).
As in \cite{Pir07} the $\beta$ value is set to 2; in this case $q=0.33$.

The cores without heating sources are most probably isothermal ($q=0$).
The dust temperatures for the cores with IRAS sources \cite{Pir07} and
the $\alpha_{i}$ values for inner regions of the cores, calculated
using these temperatures, are given in Table~\ref{powlaw}.
These values lie in the range $1.1-2.2$ in accordance with the ranges
found by other authors for the samples of high-mass star-forming
regions \cite{Muel02,Beut02,HT03,Wil05}.
The mean value of $\alpha_{i}$ is 1.6(0.3).
The values of $\langle\alpha_{i}\rangle$ for the cores
with and without IRAS sources are close to each other within
the standard deviations given in brackets.
They are 1.5(0.3) and 1.8(0.3), respectively.
The $\alpha_{out}$ index for the outer regions of the cores
is systematically higher than $\alpha_{i}$.

\section{The analysis with hydrostatic models}
\label{hydro}

As far as absolute values of power-law indices of radial
density profiles in the cores are higher on periphery than
in inner regions (see Section~\ref{powerlaw}), it is possible to use
the models of hydrostatic equilibrium spheres
confined by external pressure for their description.
The power-law index $\alpha$
of the $\rho\propto r^{-\alpha}$ dependence
rises with the distance from the center in these models.
The molecular line widths observed in the sample cores are considerably
higher that the thermal ones \cite{Pir07,Pir03}, therefore, instead of
using standard Bonnor-Ebert model one should consider a ``modified" model
(see e.g. \cite{Lai03}) which includes, apart from the thermal pressure,
the pressure from random non-thermal motions.
If the latter is homogeneous and isotropic (microturbulent approximation),
its contribution can be taken into account as an additional term
in the equation of state:

\begin{equation}
P=\rho\,(\frac{k\,T}{m}+V_{\rm nt}^2)\approx \rho\,V_{\rm nt}^2
\hspace{5mm} (V_{\rm nt}^2\gg \frac{k\,T}{m})\,,
\end{equation}

\noindent
where $m$ is a mean molecular mass,
$V_{\rm nt}$ is a mean velocity of non-thermal motions.
Density distribution can be obtained using Lane-Emden equation
for isothermal sphere:

\begin{equation}
\frac{1}{\xi^2}\frac{d}{d\xi}\Bigl (\xi^2\frac{d\,\psi}{d\xi} \Bigr )=e^{-\psi}\,,
\label{Lemden1}
\end{equation}

\noindent
where $\psi(\xi)=-\ln(\rho/\rho_c)$,
$\xi=r\cdot a$ is a dimensionless radius,
$a=V_{\rm nt}^{-1}\,\sqrt{4\pi\,G\,\rho_c}$,
where $G$ is the gravitational constant.
The boundary conditions $\psi(0)=0,\,\psi'(0)=0$ are used
for solving the equation (\ref{Lemden1}).

By solving the equation (\ref{Lemden1}) we obtain density and column density
profiles.
Having calculated the convolution of column density distribution
with the telescope beam and having fitted it into the maps, it is possible
to estimate discrepancy according to which increments of the parameters
are calculated.
An amplitude of the fitted function, an additive constant term, coordinates
of the center, $a$-parameter and radius of a sphere ($R_{\rm max}$)
are varied for each core when model profiles are fitted into the maps.
Using the values of amplitude and $a$-parameter one can calculate
$V_{\rm nt}$ and $\rho_c$.
This model, thus, allows to determine such physical parameters as
central density, dispersion of turbulent velocities and size of a core.
The comparison of the fitting results of the modified Bonnor-Ebert model
and composite power-law function (Section~\ref{powerlaw}) has not revealed
the preference of any of them according to the value of discrepancy.

The results obtained from applying the modified Bonnor-Ebert model
are given in Table~\ref{bonebert}.
Dust temperatures which are need to calculate the Plank function have been taken from
\cite{Pir07} (Table 2) for the cores with IRAS sources or have been set to 20~K
for the cores without inner sources.
The values of central density ($n_c=\rho_c/m$) and corresponding
Doppler line width ($\Delta V_{\rm nt}=\sqrt{8\ln{2}}\,V_{\rm nt}$)
are given in the columns 2 and 3 of Table~\ref{bonebert}.
Uncertainties calculated from fitting are given in brackets.
Actual uncertainties are most probably higher than those given in the table
due to uncertainty in dust absorption coefficient and uncertainty
in dust temperature for the cores without inner sources.
The gas-to-dust mass ratio is set to 100 while dust absorption coefficient
at 250~GHz is set to 1~cm$^2$~g$^{-1}$ \cite{OH94}.
An exact value of this coefficient for our sample cores is unknown.
Depending on the evolutionary phase of a core, the absorption
coefficient value can vary up to two times on the one or the other side \cite{Mot98}.
This uncertainty can lead to the uncertainty in $n_c$ up to two times and
to the uncertainty in $V_{\rm nt}$ up to $\sqrt{2}$ times.
The N$_2$H$^+$(1--0) and CS(5--4)
line widths \cite{Pir07, Pir03} observed towards the core centers are given
in Table~\ref{bonebert} for comparison.
In last two columns the values of $R_{\rm max}$ and
$\xi_{\rm max}=\xi(R_{\rm max})$ are given.
The cores with $\xi_{\rm max}>6.5$ are unstable with respect
to gravitational collapse \cite{Bon56,McCrea57}.

The distributions derived from the polytrope and the logotrope equations of state
have been fitted into the maps of several objects
obtained with high signal-to-noise ratios.
The Lane-Emden equation for the case of polytrope for $N\neq -1$
can be written in the form \cite{MH99}:

\begin{equation}
\frac{1}{\xi^2}\frac{d}{d\xi}\Bigl (\xi^2\frac{d\,\theta}{d\xi} \Bigr )=
-\frac{|N+1|}{N+1}\theta^N\,,
\label{Lemden2}
\end{equation}

\noindent
where
$\xi=r\cdot a$, $a=V_{\rm nt}^{-1}\,\sqrt{4\pi\,G\,\rho_c\,(N+1)^{-1}}$.

For the case of logotrope this equation is written in the form:

\begin{equation}
\frac{1}{\xi^2}\frac{d}{d\xi}\Bigl (\xi^2\frac{d\,\theta}{d\xi} \Bigr )=
\theta^{-1}\,,
\label{Lemden3}
\end{equation}

\noindent
where
$\xi=r\cdot a$, $a=V_{\rm nt}^{-1}\,\sqrt{20\pi\,G\,\rho_c}$.
For solving the equations (\ref{Lemden2}),(\ref{Lemden3})
the boundary conditions: $\theta(0)=1,\,\theta'(0)=0$ are used.

\begin{figure}[t!]
\setcaptionmargin{5mm}
\onelinecaptionsfalse
\centering \includegraphics[width=15cm]{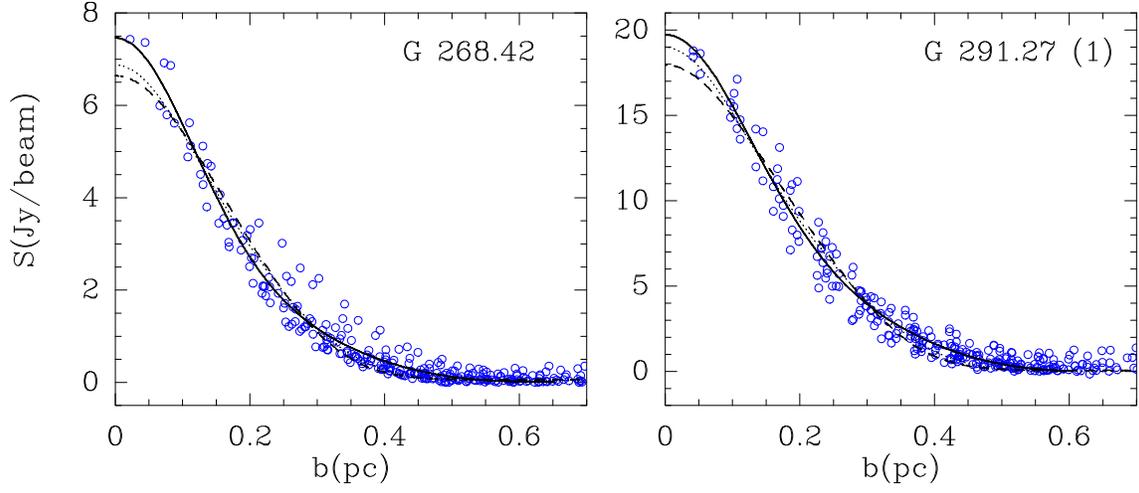}
\captionstyle{flushleft}
\caption{
The results of fitting of model distributions into the maps of two cores
with highest signal-to-noise ratios.
Solid curves correspond to the modified Bonnor-Ebert model,
the dashed ones correspond to the logotrope model,
the dotted ones correspond to the polytrope model with $N=-2$.
}
\label{fits}
\end{figure}

The polytrope models both with negative
and positive values of the $N$ index are used for the data analysis.
The comparison with the modified Bonnor-Ebert model showed that the latter
is preferable than the logotrope model in all cases and than
the polytrope models in most cases as it gives lower discrepancy values, yet,
in some cases differences are insignificant.
In two cases (G~270.26 ¨ G~285.26) an application of the polytrope models
with positive $N$ gives somewhat lower discrepancy value than
the Bonnor-Ebert model, yet, the differences are too small to confirm
reliably the preference of any model. Moreover, these cores are probably
not completely resolved in the observations (see Section~\ref{discussion}).
The physical parameters derived from the polytrope model for these two objects
are given in Table~\ref{polytrop}.
To make a final conclusion about the preference of the polytrope models
with positive $N$ one needs observations with higher signal-to-noise ratios.
The intensity distributions of dust emission and the curves corresponding
to different models are shown in Fig.~\ref{fits} as one-dimensional dependences
on the distance from the center for two representative cores.

\section{Discussion}
\label{discussion}

Taking into account probable uncertainties of the estimates derived from
fitting model distributions into the maps one can make a conclusion
that the $\Delta V_{\rm nt}$ values correspond to molecular line widths
(see Table~\ref{bonebert}) with exception of G~291.27(1) and,
probably, G~316.77(5).
In several cases the CS(5--4) line widths are considerably higher than
the N$_2$H$^+$(1--0) ones which cannot apparently be associated
with the difference in optical depth \cite{Pir07} but can reflect the difference
in non-thermal velocities of gas that effectively emits in each of these lines.
It is known that line widths of nitrogen-bearing molecules, such as
NH$_3$ are N$_2$H$^+$, are systematically lower than of the other dense gas
tracers, such as CS, HCN, HCO$^+$ (see, e.g. \cite{Pir99}),
which can be connected with chemical differentiation effects.
In our case the fact that the $\Delta V_{\rm nt}$ value is close to
the one or another molecular line width can indicate which molecular distribution
better correlates with total gas and dust distribution in given core.

The comparison of central densities obtained from the Bonnor-Ebert model
with densities derived from LVG analysis of the CS(5--4) and CS(2--1)
intensities for seven sample cores \cite{Pir07} reveals both proximity
(G269.11(1) and G294.97(1)) and discrepancy in these values.
Note, that model density estimates depend on the CS line intensity
ratios on the whole and only weakly depend
on intensities itself in the range of kinetic temperatures, densities
and column densities typical for the studied objects.
In the case of steep density gradients the CS(5--4) and CS(2--1)
emission region sizes are different, which can cause differences
in density estimates obtained with different spatial resolution.
Such gradients probably exist in G270.26 and G285.26 for which
densities according to the Bonnor-Ebert model are more than an order
of magnitude higher than densities obtained from LVG analysis of the data
convolved to the 50$''$ beam \cite{Pir07}.
The sizes of these cores in continuum are close to the main beam size
of the antenna pattern and these sources are probably not completely resolved
even in continuum observations.
Note, that at high densities and low temperatures the CS abundances can drop
due to depletion onto dust grains toward the centers of the cores without inner sources.
This effect can underestimate central densities calculated from LVG
analysis.
However, it is probably not the case for the considered objects which
contain inner heating sources.
The LVG densities for G268.42 and G291.27(1) appear to be
$\sim 2$ times lower than those derived from the Bonnor-Ebert model
which can be connected with the fact that the actual value
of dust absorption coefficient is higher than the standard value
used for these sources (see Section~\ref{hydro}).
The density obtained from LVG calculations for G265.14(1)
is an order of magnitude higher than the derived from the Bonnor-Ebert model.

It is important to note that the cores associated with high-mass
star-forming regions can be inhomogeneous on the scales unresolved
by modern instruments (see \cite{PZ08} and references therein).
There are implications that enhanced density regions
in the sample cores also could not fill the telescope beam \cite{Pir07}.
It is possible that an excess of the LVG density upon the density derived
from the Bonnor-Ebert model for G265.14(1) is connected with the fact that
the CS emission comes from the clumps with enhanced density and
low filling factor whereas density profile derived from dust emission
is related to the values averaged over the telescope beam.

In general, the cores studied in the present paper appear to look very like
hydrostatic Bonnor-Ebert spheres taking into account turbulent pressure,
while such physical parameters as central density,
velocity dispersion and size of a core calculated from this model do not
contradict independent estimates.
Note, however, that the $\xi_{\rm max}$ value is higher than the critical
one more than in half cases, implying their instability in the framework
of the Bonnor-Ebert model.
We found no correlation between a presence or an absence of closely located
IRAS source and stability of a core.
Although there are indications in literature that the Bonnor-Ebert profiles
can also hold for unstable slowly collapsing objects \cite{M05,Kan05},
this fact needs further exploration.
Note, that model calculations of the structure of turbulent molecular clouds
predict an existence of non-equilibrium density fluctuations
which density profiles can be close to those followed
from the Bonnor-Ebert model, while physical parameters derived from this
model considerably differ from those initially given \cite{BP03}.
In order to get definite conclusion is it correct to apply hydrostatic
models to the considered objects the observational data with higher
signal-to-noise ratios are needed.

\section{Conclusions}

The estimates of radial density profiles for the sample of dense cores
associated with high-mass star-forming regions from southern hemisphere
have been done using the data of observations in continuum at 250~GHz.
For this purpose two-dimensional functions are fitted into the observed maps.
The function fitted is a convolution of the telescope beam function and
dust column density distribution which is set either arbitrary power-law
function or the one followed from various hydrostatic models.

An analysis using power-law functions showed that radial density profiles
for the inner regions of 16 cores (at distances $\la 0.2-0.8$~pc from the center)
are close on average to the $\rho\propto r^{-\alpha}$ dependence,
where $\alpha=1.6\pm 0.3$.
In the outer regions density drops steeply.

Among hydrostatic models the modified Bonnor-Ebert model
which describes a sphere with thermal and non-thermal (microturbulent)
motions confined by external pressure, as well as the polytrope
and the logotrope models have been considered.
The modified Bonnor-Ebert model gives the best results practically in all cases.
With a help of this model the estimates of central density,
non-thermal velocity dispersion and core size have been done.
The calculated values of central densities lie in the range:
$6\cdot 10^4-2\cdot 10^7$~cm$^{-3}$,
non-thermal velocity dispersions expressed in the form of Doppler line widths
lie in the range: $2-12$~km/s, core sizes are $0.24-1.05$~pc.
The comparison of the central densities with the densities obtained from
calculations of the CS molecule excitation \cite{Pir07} has been done.
The differences in the values revealed in several cases are
probably connected with the presence of density inhomogenities on
the scales smaller than the telescope beam.
The non-thermal velocity dispersions are in agreement with the values
obtained from molecular line observations in most cases.

\begin{acknowledgments}
This work has been carried out with a support from RFBR
(06-02-16317, 08-02-00628 grants) and from the program
``Active Processes and Stochastic Structures in the Universe" 
of the Physical Sciences Division of RAS.
\end{acknowledgments}

{}

\newpage

\newpage

\newpage

\newpage

\begin{table}[hbtp]
\setcaptionmargin{0mm}
\onelinecaptionstrue
\captionstyle{flushleft}
\caption[]{Source list}
\bigskip
\begin{tabular}{cccc}
\noalign{\hrule}\noalign{\smallskip}
Source    & RA (2000)                      & Dec (2000)                 & $D$   \\
          & ${\rm (^h)\  (^m)\  (^s)\ }$ &($^o$ $^{\prime}$ $^{\prime\prime}$) & (kpc) \\
\noalign{\smallskip}\hline\noalign{\smallskip}
G~265.14$+$1.45  &08 59 24.7   &$-$43 45 22  & ~1.7 \cite{Zin95} \\
G~268.42$-$0.85  &09 01 54.3   &$-$47 43 59  & ~1.3 \cite{Zin95} \\
G~269.11$-$1.12  &09 03 32.8   &$-$48 28 39  & ~2.6 \cite{Zin95} \\
G~270.26$+$0.83  &09 16 43.3   &$-$47 56 36  & ~2.6 \cite{Zin95} \\
G~285.26$-$0.05  &10 31 30.0   &$-$58 02 07  & ~4.7 \cite{Zin95} \\
G 291.27$-$0.71  &11 11 49.9   &$-$61 18 14  & ~2.7 \cite{BB93} \\
G~294.97$-$1.73  &11 39 12.6   &$-$63 28 47  & ~1.2 \cite{Zin95} \\
G~316.77$-$0.02  &14 44 58.9   &$-$59 48 29  & ~3.1 \cite{Juv96} \\
G~345.01$+$1.80  &16 56 45.3   &$-$40 14 03  & ~2.1 \cite{Juv96} \\
G~351.41$+$0.64  &17 20 53.4   &$-$35 47 00  & ~1.7 \cite{Nec78} \\
\noalign{\smallskip}\hline\noalign{\smallskip}
\end{tabular}

\label{list}
\end{table}

\newpage

\begin{table}[htbp]
\setcaptionmargin{0mm}
\onelinecaptionstrue
\captionstyle{flushleft}
\caption[]{The results obtained from fitting by power-law functions}
\bigskip
\begin{tabular}{lccccccc}
\noalign{\hrule}\noalign{\smallskip}
Source  & $T$ & $p_{i}$ & $p_{out}$ & $\alpha_{i}$ &$\Delta \Theta_{\rm S}$ & $R_{\rm S}$ & IRAS\\
         & (K) &         &           &            & ($^{\prime\prime}$)   &(pc) \\
\noalign{\smallskip}\hline\noalign{\smallskip}

G~265.14$+$1.45 (1)  & 30 & 0.54(0.07) & 1.3(0.1)   & 1.14   & 25.0 & 0.21 & +  \\
G~265.14$+$1.45 (2)  &    & 0.33(0.05) & 0.89(0.08) & 1.33   & 37.5 & 0.31 & -- \\
G~265.14$+$1.45 (3)  & 30 & 0.79(0.05) & 1.4(0.1)   & 1.39   & 32.5 & 0.27 & +  \\
G~268.42$-$0.85      & 35 & 1.04(0.02) & 1.42(0.06) & 1.65   & 40.0 & 0.25 & +  \\
G~269.11$-$1.12 (1)  &    & 1.05(0.04) & 1.60(0.08) & 2.05   & 27.5 & 0.35 & -- \\
G~270.26$+$0.83      & 29 & 1.09(0.09) & 1.52(0.05) & 1.69   & 17.5 & 0.22 & +  \\
G~285.26$-$0.05 (1)  & 33 & 1.36(0.03) & 1.84(0.04) & 1.97   & 22.5 & 0.51 & +  \\
G~291.27$-$0.71 (1)  & 25 & 0.83(0.01) & 1.24(0.02) & 1.42   & 37.5 & 0.49 & +  \\
G~291.27$-$0.71 (3)  &    & 0.95(0.06) & 1.2(0.3)   & 1.95   & 60.0 & 0.79 & -- \\
G~291.27$-$0.71 (4)  &    & 0.50(0.05) & 1.5(0.5)   & 1.50   & 32.5 & 0.43 & -- \\
G~294.97$-$1.73 (1)  & 27 & 0.72(0.03) & 0.97(0.04) & 1.31   & 35.0 & 0.20 & +  \\
G~294.97$-$1.73 (2)  &    & 0.81(0.06) & 0.94(0.05) & 1.81   & 30.0 & 0.17 & -- \\
G~316.77$-$0.02 (4)  &    & 0.92(0.17) & 0.92(0.17) & 1.92   &      &      & -- \\
G~316.77$-$0.02 (5)  &    & 1.16(0.26) & 1.16(0.26) & 2.16   &      &      & -- \\
G~345.01$+$1.80 (2)  &    & 0.56(0.13) & 0.56(0.13) & 1.56   &      &      & -- \\
G~351.41$+$0.64 (2)  &    & 0.44(0.07) & 1.05(0.08) & 1.44   & 25.0 & 0.21 & -- \\
\noalign{\smallskip}\hline\noalign{\smallskip}
\end{tabular}

\footnotesize
\begin{flushleft}
$T$(K) is a dust temperature \cite{Pir07},
$p_{i}$ and $p_{out}$ are power-law indices of the fitting functions
for inner and outer core regions, respectively,
$\alpha_{i}$ is a power-law index of radial density profile for inner regions,
$\Delta \Theta_{\rm S}$ and $R_{\rm S}$ are angular and linear
separation radius values between inner and outer regions, respectively.
\end{flushleft}
\normalsize

\label{powlaw}
\end{table}

\newpage

\begin{table}[hbtp]
\setcaptionmargin{0mm}
\onelinecaptionstrue
\captionstyle{flushleft}
\caption[]{The results of using the modified Bonnor-Ebert model}
\bigskip
\begin{tabular}{lcrccccc}
\noalign{\hrule}\noalign{\smallskip}
Source  & $n_c$    &$\Delta V_{\rm nt}$ &$\Delta V$(N$_2$H$^+$) &$\Delta V$(CS) & $R_{\rm max}$ & $\xi_{\rm max}$ & IRAS \\
         & (cm$^{-3}$) &  (km/s)            &  (km/s)               &  (km/s)       &    (pc)       \\
\noalign{\smallskip}\hline\noalign{\smallskip}

G~265.14$+$1.45 (1)    & 1.4(0.4) 10$^5$ & 3.6(1.3) & 2.58(0.04) & 2.6(0.1) & 0.24(0.02) &   3.3(1.2)  & +   \\
G~265.14$+$1.45 (2)    & 1.6(0.4) 10$^5$ & 2.7(0.7) & 2.34(0.03) & 2.0(0.1) & 0.27(0.02) &   5.2(1.1)  & --  \\
G~265.14$+$1.45 (3)    & 5.7(0.9) 10$^4$ & 4.3(1.2) & 1.95(0.03) & 4.3(1.2) & 0.36(0.02) &   2.7(0.7)  & +   \\
G~268.42$-$0.85        & 1.9(0.2) 10$^6$ & 3.9(0.2) & 2.64(0.09) & 3.4(0.1) & 0.36(0.01) &  16.4(0.8)  & +   \\
G~269.11$-$1.12 (1)    & 1.9(0.5) 10$^6$ & 5.2(0.9) & 3.23(0.03) & 5.8(0.1) & 0.52(0.02) &  18.4(2.4)  & --  \\
G~270.26$+$0.83        & 1.6(1.4) 10$^7$ & 4.6(2.7) & 3.42(0.03) & 4.2(0.1) & 0.36(0.01) &  41.7(16.6) & +   \\
G~285.26$-$0.05 (1)    & 2.4(0.6) 10$^7$ & 6.4(1.0) & 3.09(0.39) & 5.0(0.1) & 0.67(0.02) &  66.3(7.4)  & +   \\
G~291.27$-$0.71 (1)    & 2.1(0.1) 10$^6$ & 11.7(0.4)& 2.31(0.15) & 5.0(0.2) & 0.76(0.01) &  12.2(0.4)  & +   \\
G~291.27$-$0.71 (3)    & 1.2(0.2) 10$^5$ & 6.7(1.5) &            &          & 0.50(0.02) &   3.3(0.7)  & --  \\
G~291.27$-$0.71 (4)    & 2.1(0.2) 10$^5$ & 4.1(0.4) & 2.29(0.03) & 2.9(0.1) & 1.05(0.09) &  15.4(1.7)  & --  \\
G~294.97$-$1.73 (1)    & 3.2(0.3) 10$^5$ & 2.2(0.1) & 2.36(0.04) & 2.9(0.3) & 0.35(0.01) &  12.0(0.8)  & +   \\
G~294.97$-$1.73 (2)    & 6.2(1.0) 10$^5$ & 2.2(0.3) & 2.57(0.06) & 2.3(0.2) & 0.44(0.03) &  21.0(2.3)  & --  \\
G~316.77$-$0.02 (4)    & 1.9(0.5) 10$^5$ & 3.4(0.6) & 3.31(0.01) & 4.0(0.3) & 0.80(0.07) &  13.6(2.2)  & --  \\
G~316.77$-$0.02 (5)    & 6.2(1.2) 10$^4$ & 6.5(2.1) & 3.20(0.01) & 3.9(0.3) & 0.53(0.03) &   2.7(0.9)  & --  \\
G~345.01$+$1.80 (2)    & 3.4(2.0) 10$^5$ & 3.3(1.5) & 3.30(0.01) & 3.1(0.1) & 0.45(0.09) &  10.3(3.9)  & --  \\
G~351.41$+$0.64 (2)    & 1.4(0.3) 10$^6$ & 8.0(1.8) & 4.38(0.02) & 7.7(0.1) & 0.28(0.02) &   5.4(1.1)  & --  \\
\noalign{\smallskip}\hline\noalign{\smallskip}
\end{tabular}

\footnotesize
\begin{flushleft}
$n_c$ is a central density,
$\Delta V_{\rm nt}$ is a Doppler width corresponding to
mean velocity of non-thermal motions,
$R_{\rm max}$ is a sphere radius,
$\xi_{\rm max}$ is a dimensionless radius.
\end{flushleft}
\normalsize

\label{bonebert}
\end{table}

\begin{table}[hbtp]
\setcaptionmargin{0mm}
\onelinecaptionstrue
\captionstyle{flushleft}
\caption[]{The results of using the polytrope model}
\bigskip
\begin{tabular}{lrcrc}
\noalign{\hrule}\noalign{\smallskip}
Source  & $N$ & $n_c$       &$\Delta V_{\rm nt}$ & $R_{\rm max}$ \\
         &     & (cm$^{-3}$) &  (km/s)            &    (pc)      \\
\noalign{\smallskip}\hline\noalign{\smallskip}

G~270.26$+$0.83        &  8 & 3.4(0.5) 10$^6$ & 4.8(0.5) & 0.58(0.07) \\
G~285.26$-$0.05 (1)    & 15 & 2.5(0.5) 10$^7$ & 7.6(1.0) & 0.82(0.02) \\
\noalign{\smallskip}\hline\noalign{\smallskip}
\end{tabular}

\label{polytrop}
\end{table}

\end{document}